\documentclass[pra,aps,eqsecnum,amsmath,amssymb,showpacs,showkeys]{revtex4}
\usepackage{graphicx}
\newcommand{\qp}{\mathsf{P}}
\newcommand{\mm}{\mathsf{M}}
\newcommand{\X}{\mathsf{X}}
\newcommand{\ip}{\mathsf{\Pi}}
\newcommand{\U}{\mathsf{U}}
\newcommand{\wsi}{\widetilde{\Psi}}
\newcommand{\wpi}{\widetilde{\Phi}}
\newcommand{\wpp}{\widetilde{p}}
\newcommand{\wqq}{\widetilde{q}}
\newcommand{\tr}{\rm{tr}}

\begin{document}

\preprint{}

\title{{\bf No-cloning theorem for a single POVM}}

\author{Alexey E. Rastegin}
 \affiliation{Department of Theoretical Physics, Irkutsk State University,
Gagarin Bv. 20, Irkutsk 664003, Russia}
 \email{rast@api.isu.ru}

\begin{abstract}
Cloning of statistics of general
quantum measurement is discussed. The presented approach is
connected with the known concept of observable cloning, but
differs in some essential respects. The reasons are illustrated
within some variety of the B92 protocol. As it is shown, there
exist pairs of states such that the perfect cloning of given POVM
is not possible. We discuss some properties of these intolerant
sets. An example allowing the perfect cloning is presented as
well.
\end{abstract}



\maketitle

\section{Introduction}

In the quantum world, powerful tools for
information processing can be found \cite{nielsen}. The principal
idea is encoding information into quantum states. Due to quantum
lability, a copying problem becomes very important. The main
point in this regard is expressed by the no-cloning theorem
\cite{wootters}. The result was extended to the case of mixed
states \cite{barnum}. Surprisingly, in composite systems the
no-cloning principle for orthogonal states holds \cite{mor}. An
approximate cloning of quantum states was firstly studied by
Bu\v{z}ek and Hillery \cite{hillery1}. Various results and
scenarios have been developed in this trend \cite{fanh}. The
problem of mixed-state cloning remains attractive
\cite{ekert,liu}. Quantum cloning is naturally connected with
analysis of eavesdropping strategies and Bell's inequality
\cite{gisin,zbinden}. In the context of quantum cryptography, the
asymmetric cloning machines are interesting \cite{rast02,filip}.
Quantum cloning can be used for joint measurement of noncommuting
observables \cite{ariano}. In view of the stronger no-cloning
theorem \cite{jozsa1}, the cloning with {\it a priori} information
in the ancilla have been studied \cite{rast033,rast032}.

For real information systems, there are crucial aspects such as
the identity corroboration or the access control. These problems
are more complicated at the quantum level. In a multi-user line, a
work of any quantum repeater is essentially limited \cite{genoni}.
The bounds are posed in a form of information-disturbance
trade-off obtained for a general case \cite{banaszek,devetak} as
well as for specified scenarios \cite{sacchi,buscemi}. Some
authentication protocols have been examined
\cite{leung,curty,yen09}. The concept of observable cloning was
proposed in Refs. \cite{paris6,paris7}. In the present work, this
idea is developed from another point of view. In particular, we
address the question whether a statistics of quantum measurement
can be utilized as additional source of secrecy. Such statistical
data might be adopted in the context of multi-user network, for
instance, as a kind of ''certificate'' for access mechanism. That
is, the statistics is known not only to users but to a trusted
entity as well. So we ask a character of vulnerability of
measurement statistics under evil activity. It is assumed that the
used POVM may be known to an intruder. To provide a secrecy, more
than one sets of quantum states should be applicable to encryption
with the same technical equipment. We describe the example related
to the B92 scheme. Meantime, the presented results may be useful
in more abstract sense and other contexts.

\section{Description of the problem}

As a part of the communication, given quantum channel
can be used by legitimate users for various aims. In this work we
mean the following scenario. Physically, the sender (Alice)
secretly chooses a state from the specified set ${\cal{B}}$ and
sends this state to the receiver (Bob). Then Bob performs some
quantum measurement ${\mathcal{M}}$ generally described by
''positive operator-valued measure'' (see Fig. 1). Recall that
POVM ${\mathcal{M}}=\{{\mathsf{M}}_m\}$ is a set of positive
operators obeying $\sum_m{\mathsf{M}}_m={\mathbf{1}}$
\cite{nielsen}. The utilized quantum measurement is assumed to be
known to both parties. Meanwhile, Alice may principally choose the
set ${\cal{B}}$ without giving Bob a notice which is required for
quantum key distribution too. The intruder (Trudy) entangles probe
`T' with quantum carrier `A' and further acts with respect to own
pursued interests. Of course, the actually sent state is unknown
for Trudy. We now describe an example where Trudy's prior knowledge of used
POVM does not imply her knowledge of actual set ${\mathcal{B}}$.

\begin{figure}
\centering
\includegraphics[width=8.2cm]{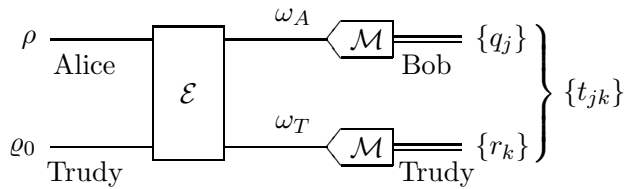}
\caption{Two input qudits are processed by Trudy
via channel $\cal{E}$. Then she sends qudit `A' to Bob. Finally,
Bob and Trudy perform POVM measurement ${\mathcal{M}}$, every on
own qudit.}
\end{figure}

In the B92 scheme \cite{bennett}, Alice encodes bits 0 and 1 into
nonorthogonal states $|\eta_{+}\rangle\neq|\eta_{-}\rangle$. In
its original version, Bob randomly measures one of two projectors
$|\theta_{+}\rangle\langle\theta_{+}|$ and
$|\theta_{-}\rangle\langle\theta_{-}|$, where
$|\theta_{i}\rangle\perp|\eta_{i}\rangle$. From the viewpoint of
implementation, the measurement is realized by carrying out with
probability $1/2$ one of two PVMs
$\{|\eta_{+}\rangle\langle\eta_{+}|,|\theta_{+}\rangle\langle\theta_{+}|\}$
and
$\{|\eta_{-}\rangle\langle\eta_{-}|,|\theta_{-}\rangle\langle\theta_{-}|\}$.
Formally, this is described by the four-element POVM
\begin{equation}
{\mathcal{N}}:=\left\{\frac{1}{2}|\eta_{+}\rangle\langle\eta_{+}|,\frac{1}{2}|\theta_{+}\rangle\langle\theta_{+}|,
\frac{1}{2}|\eta_{-}\rangle\langle\eta_{-}|,\frac{1}{2}|\theta_{-}\rangle\langle\theta_{-}|\right\}
\ . \label{fepovm}
\end{equation}
Further, Bob combines outcomes `$\eta_{\pm}$' into one
inconclusive answer. More economical way uses the optimal POVM for
unambiguous discrimination between $|\eta_{+}\rangle$ and
$|\eta_{-}\rangle$ \cite{palma}. However, in this case only one
prescribed pair of states can be utilized. On the other hand,
Alice and Bob can take any two nonorthogonal states from the set
$\{|\eta_{\pm}\rangle,|\theta_{\pm}\rangle\}$ with the same
equipment for realizing the POVM ${\mathcal{N}}$. The legitimate
users can adopt four acceptable pairs for different s\'{e}ances of
quantum key distribution. This is not possible with the optimal
unambiguous discrimination. By numerical tests for two-state
protocol, it was found that optimal intruder's probe needs only
two dimensions \cite{fuchs}. But such a conclusion is hardly valid
when one involves all states from the set
$\{|\eta_{\pm}\rangle,|\theta_{\pm}\rangle\}$. A discussion of
this question would take us to far afield. We merely note that
a potential resource may be dealt within the B92 scheme.

By ${\mathbb{C}}^d$ we denote $d$-dimensional complex vector
space, i.e. the state space of $d$-level system (qudit). After
interaction the system `AT' is in a state
$\Omega={\cal{E}}(\rho\otimes\varrho_0)$, where $\varrho_0$ is the
initial state of `T' assumed to be fixed. In general, the
evolution of open quantum system is described by a completely
positive trace-preserving linear map \cite{nielsen} called
''quantum channel.'' The output density matrices are expressed as
partial traces $\omega_A={\tr}_T(\Omega)$ and
$\omega_T={\tr}_A(\Omega)$ over the corresponding subspaces. In
Bob's measurement on qudit `A', $j$-th outcome occurs with the
probability
$q_j:={\tr}\bigl(({\mm}_j\otimes{\mathbf{1}})\Omega\bigr)={\tr}_A({\mm}_j{\,}\omega_A)$.
If Trudy try to copy the original statistics then she performs the
measurement ${\mathcal{M}}$ on qudit `T' with the probability of
$k$-th outcome equal to
$r_k:={\tr}\bigl(({\mathbf{1}}\otimes{\mm}_k)\Omega\bigr)={\tr}_T({\mm}_k{\,}\omega_T)$.
In the case considered, each of the two probability distributions
$\{q_j\}$ and $\{r_k\}$ is actually marginal of the joint
distribution $\{t_{jk}\}$ given by
$t_{jk}:={\tr}\bigl(({\mm}_j\otimes{\mm}_k)\Omega\bigr)$. Both the
probability distributions $\{q_j\}$ and $\{r_k\}$ should be
compared with the original distribution
$\bigr\{p_i:={\tr}_A({\mm}_i\rho)\bigl\}$. In addition, Trudy
would like to conceal her activity. One of possible approach is to
define the perfect standard by
\begin{equation}
\sum\nolimits_k t_{jk}=p_j \ , \quad \sum\nolimits_j t_{jk}=p_k
\ . \label{broad}
\end{equation}
We will say ''broadcasting'' of statistics, when results of
copying process are assumed to be compared just with the standard
(\ref{broad}). This wording concurs with the notation
emerged in Ref. \cite{barnum}. If Trudy means a replication of
POVM statistics as such, then broadcasting can be utilized.
However, this way is rather insufficient when Trudy intends to use
the obtained results in future action. For example, her data
together with Bob's data may be exposed to a trusted authority in
some stage. Here she has to consider a special form of
broadcasting in which the perfect standard is expressed as
$t_{jk}=p_{j}p_{k}$ for all $j$ and $k$. That is, the
factorization of joint distribution is wanted as well. This
process will be called ''cloning'' of statistics. As is
mentioned above, statistics cloning may have a practical sense. It
is also interesting from the viewpoint of fundamental limitations on
manipulation with quantum information.

After choice of the standard for comparison, we should adopt good
figure of merit. In the context of statistics cloning, the actual
joint distribution $\{t_{jk}\}$ is compared with the perfect
standard $\{p_{j}p_{k}\}$. The notion of relative entropy is very
useful in many respects \cite{nielsen}. Recall that the relative
entropy of $\{p_j\}$ to $\{q_j\}$ is defined by
$H(p_j||q_j):=\sum_j p_j\ln({p_j}/{q_j})$. If given POVM
$\{{\mathsf{M}}_m\}$ has been cloned perfectly then
$H(t_{jk}||p_jp_k)=0$. Otherwise, a cloning process is
approximate. Let ${\cal{B}}=\{\rho_{\mu}\}$ be a set of density
operators on ${\mathbb{C}}^d$. Then a merit of cloning can
naturally be evaluated by the measure
$H_{\cal{B}}:=\sup\{H(t_{jk}||p_jp_k):{\,}\rho\in{\cal{B}}\}$.

To each observable we can assign a ''projector-valued measure''
(PVM). In this sense, the concept of observable cloning is
involved into the above reasons. On the other hand, the authors of
Ref. \cite{paris6} expressed cloning of an observable in terms of
its mean value. Another point is that they focused an attention
on the incompatibility of observables. At the same time, a
structure of input quantum states can be crucial for use of
cloning in eavesdropping process \cite{mor}. In effect, a small
number of states is typically used in the protocols of quantum
cryptography. Further, our reasons are expressed purely in terms
of probability distributions. Note that there are other ways to
evaluate a merit of POVM cloning. For examle, we would adopt the
trace distance which has clear operational meaning. In general, a
good choice for figure of merit may be specified by the actual
context.

\section{No-cloning of a single POVM}

In this section the main result will be presented. For
any operator ${\mathsf{X}}$ on ${\mathbb{C}}^d$, the operator
${\X}^{\dagger}{\X}$ is positive. The operator $|{\mathsf{X}}|$ is
defined as a unique positive square root of
${\mathsf{X}}^{\dagger}{\mathsf{X}}$. The eigenvalues of
$|{\mathsf{X}}|$ counted with their multiplicities are {\it
singular values} $s_j({\mathsf{X}})$ of operator ${\mathsf{X}}$
\cite{watrous1}. So the fidelity between two density operators
$\rho$ and $\omega$ is defined by \cite{nielsen,uhlmann76}
\begin{equation}
F_0(\rho,\omega)=\sum\nolimits_{j=1}^d s_j(\sqrt{\rho}\sqrt{\omega})
\equiv{\tr}|\sqrt{\rho}\sqrt{\omega}|
\ . \label{defin11}
\end{equation}
Note that Jozsa \cite{jozsa94} used the word ''fidelity'' for the
square of the right-hand side of (\ref{defin11}). This may be more
convenient sometimes \cite{rast032,rast031,holevo}. However, we
will further use the definition (\ref{defin11}). The fidelity
function enjoys many useful properties including the quantum-classical
relation \cite{caves}. The classical fidelity between probability
distributions $\{p_j\}$ and $\{q_j\}$ is given by
\begin{equation}
{\cal{F}}(p_j,q_j):=\sum\nolimits_j \sqrt{p_j q_j} \ .
\label{fidcl}
\end{equation}
The authors of Ref. \cite{caves} showed that the fidelity satisfies
\begin{equation}
F(\rho,\omega)=\min {\cal{F}}(p_m,q_m)
\ , \label{relfid0}
\end{equation}
where $p_m={\tr}_A({\mathsf{M}}_m\rho)$,
$q_m={\tr}_A({\mathsf{M}}_m\omega)$ and the minimization is over
all POVMs. The concept of purification is also needed. Adding
another qudit `B', we re-express mixed states $\rho$ and $\omega$
of qudit `A' as partial traces
$\rho={\rm{tr}}_{B}(|\Psi\rangle\langle\Psi|)$ and
$\omega={\rm{tr}}_{B}(|\Phi\rangle\langle\Phi|)$. Here pure states
$|\Psi\rangle$ and $|\Phi\rangle$ of the total system `AB' are
purifications of $\rho$ and $\omega$. Then we have \cite{jozsa94}
\begin{equation}
F(\rho,\omega)=
\max\left|\langle\Psi|\Phi\rangle\right|
\ , \label{pr4}
\end{equation}
where the maximum is taken over all purifications $|\Psi\rangle$
of $\rho$ and $|\Phi\rangle$ of $\omega$.

{\bf Theorem 1.} {\it A POVM $\{{\mathsf{M}}_m\}$ cannot be cloned
with $H_{\cal{B}}=0$ over each pair $\{\rho,\omega\}$ of states
such that}
\begin{equation}
{\cal{F}}(p_m,q_m)^2<F(\rho,\omega)
\ . \label{the3}
\end{equation}

{\bf Proof.} Suppose that POVM $\{{\mathsf{M}}_m\}$ has been
cloned perfectly over two different inputs $\rho'$ and $\rho''$.
Then associated probabilities are $t'_{jk}=p'_jp'_k$ and
$t''_{jk}=p''_jp''_k$. For the outputs
$\Omega'\equiv{\cal{E}}(\rho'\otimes\varrho_0)$ and
$\Omega''\equiv{\cal{E}}(\rho''\otimes\varrho_0)$, the measurement
$\{{\mathsf{M}}_j\otimes{\mathsf{M}}_k\}$ gives classical fidelity
\begin{equation}
{\cal{F}}(t'_{jk},t''_{jk})=\sum\nolimits_{jk} \left(p'_jp'_kp''_jp''_k\right)^{1/2}=
\sum\nolimits_{j}(p'_jp''_j)^{1/2} \sum\nolimits_{k}(p'_kp''_k)^{1/2}=
{\cal{F}}(p'_j,p''_j)^2
\ . \label{fidsq}
\end{equation}
Due to the statistical interpretation (\ref{relfid0}), we have
$F(\Omega',\Omega'')\leq{\cal{F}}(p'_j,p''_j)^2$. On the other
hand, the fidelity cannot decrease under any trace-preserving quantum
operation \cite{nielsen}, that is
\begin{equation}
F(\rho',\rho'')=F(\rho'\otimes\varrho_0,\rho''\otimes\varrho_0)\leq F(\Omega',\Omega'')
\ , \label{nonde}
\end{equation}
where the multiplicativity is used. Together the last two relations imply that
\begin{equation}
F(\rho',\rho'')\leq {\cal{F}}(p'_j,p''_j)^2
\ . \label{cothe3}
\end{equation}
This inequality is the negation of precondition (\ref{the3}). $\blacksquare$

Thus, the inequality (\ref{cothe3}) is necessary condition for
perfect cloning of POVM over two input quantum carriers. The
statement of Theorem 1 gains novel view on cloning of measurement
statistics. Let us return to the above example with the B92 protocol.
Writing
\begin{equation}
|\eta_{\pm}\rangle=\cos\eta|0\rangle\pm\sin\eta|1\rangle
\ , \quad |\theta_{\pm}\rangle=-\sin\eta|0\rangle\pm\cos\eta|1\rangle
\ ,   \label{etpar}
\end{equation}
where $\eta\in(0;\pi/4)$, we find the distributions:
$\bigl\{1/2,0,\cos^2 2\eta{\,}/2,\sin^2 2\eta{\,}/2\bigr\}$ for
$|\eta_{+}\rangle$ and $\bigl\{\cos^2 2\eta{\,}/2,\sin^2
2\eta{\,}/2,1/2,0\bigr\}$ for $|\eta_{-}\rangle$. The classical
fidelity between them is equal to
$\cos2\eta=\langle\eta_{+}|\eta_{-}\rangle$. Since
${\cal{F}}=F\neq0,1$, the condition (\ref{the3}) is provided. By a
symmetry argument, the same conclusion holds for all the four
pairs of nonorthogonal states from the set
$\{|\eta_{\pm}\rangle,|\theta_{\pm}\rangle\}$. In each case, Trudy
is unable to reach $H_{\cal{B}}=0$ over the pair. When quantum key
distribution is not assumed, Alice can change the used pair
independently of Bob. If both the Bob's and Trudy's data are
exposed to a trusted authority then an intrusion will be detected
with high probability. Certainly, an incompatibility property is
basic in quantum theory. But another key aspect is dictated by a
set of states into which we encode information.

\section{Some properties of intolerant sets}

There are two clear cases, $F(\rho,\omega)=0$ and
$F(\rho,\omega)=1$, for which the condition (\ref{the3}) cannot be
valid independently of POVM to be cloned. So we will mean that
$0<F(\rho,\omega)<1$. In such a case, if we have the equality
\begin{equation}
{\cal{F}}(p_m,q_m)=F(\rho,\omega)
\ , \label{relfid00}
\end{equation}
then the perfect cloning of given POVM over the pair $\{\rho,\omega\}$ is
not possible. It is known that for commuting states this equality takes place if and
only if \cite{caves}
\begin{equation}
z_m{\mathsf{M}}_m^{1/2}{\rho}^{1/2}={\mathsf{M}}_m^{1/2}{\omega}^{1/2}
\label{linrel1}
\end{equation}
for all $m$ and some set $\{z_m\}$ of complex numbers. For any two
density operator, a minimized POVM can always be constructed
\cite{nielsen,caves}. On the other hand, for the given POVM we can
select those pairs of density operators that satisfy
(\ref{relfid00}). Here the transitivity takes place, since if the
pairs $\{\rho,\omega\}$ and $\{\omega,\varrho\}$ obey the
condition of the form (\ref{linrel1}) then the pair
$\{\rho,\varrho\}$ does enjoy this as well. Indeed, we
have the equalities (\ref{linrel1}) and
$\xi_m{\mathsf{M}}_m^{1/2}{\omega}^{1/2}={\mathsf{M}}_m^{1/2}{\varrho}^{1/2}$,
whence
$\xi_{m}z_m{\mathsf{M}}_m^{1/2}{\rho}^{1/2}={\mathsf{M}}_m^{1/2}{\varrho}^{1/2}$.
So we obtain some equivalence class of density operators. It turns
out that this class is uncountably infinite. In the case of PVM
$\{{\mathsf{P}}_m\}$ and pure states $\rho=|\psi\rangle\langle\psi|$, 
$\omega=|\phi\rangle\langle\phi|$, we have
$p_m=\langle\psi|{\mathsf{P}}_m|\psi\rangle$ and
$q_m=\langle\phi|{\mathsf{P}}_m|\phi\rangle$. The following lemma
is proven in Appendix A.

{\bf Lemma 2.} {\it Let PVM $\{{\mathsf{P}}_m\}$ be given. For
each $|\psi\rangle\in{\mathbb{C}}^d$ and any
$f\in\bigl[{\rm{min}}\{p_m\}^{1/2};1\bigr]$ there exists pure
state $|\phi\rangle$ such that the equality (\ref{relfid00}) holds
and $|\langle\psi|\phi\rangle|=f$.}

The parameter $f$ is continuously varied in the interval from
${\rm{min}}\{p_m\}^{1/2}$ to 1. So we obtain uncountably infinite
set of pure states with the desired properties for any prescribed
$|\psi\rangle$. Using the concept of purifications, this result
can be extended to  mixed states.

{\bf Theorem 3.} {\it Let PVM $\{{\mathsf{P}}_m\}$ be given. For
each density operator $\rho$ on ${\mathbb{C}}^d$ and any
$f\in\bigl[{\rm{min}}\{p_m\}^{1/2};1\bigr]$ there exists density
operator $\omega$ such that the equality (\ref{relfid00}) holds
and $F(\rho,\omega)=f$.}

{\bf Proof.} The set $\bigl\{{\ip}_m={\mathsf{P}}_m\otimes{\mathbf{1}}\bigr\}$ is a
PVM on the space $\bigl({\mathbb{C}}^d\bigr)^{\otimes2}$ of the
qudits `A' and `B'. Let us fix some purification
$|\wsi\rangle$ of given mixed state $\rho$. To
the chosen state $|\wsi\rangle$ assign a pure state $|\wpi\rangle$
such that $|\langle\wsi|\wpi\rangle|={\cal{F}}(\wpp_m,\wqq_m)$,
where $\wpp_m=\langle\wsi|{\ip}_m|\wsi\rangle$,
$\wqq_m=\langle\wpi|{\ip}_m|\wpi\rangle$.
This possibility is ensured by Lemma 2. We now define mixed state
$\omega$ as the partial trace
$\omega={\rm{tr}}_{B}(|\wpi\rangle\langle\wpi|)$. It follows
from the properties of partial trace that
$\wpp_m={\rm{tr}}_{A}({\mathsf{P}}_m\rho)\equiv p_m$ and
$\wqq_m={\rm{tr}}_{A}({\mathsf{P}}_m\omega)\equiv q_m$ for
the prescribed $\rho$ and the built $\omega$. So, the classical fidelity
${\cal{F}}(p_m,q_m)$ satisfies
\begin{equation}
{\cal{F}}(p_m,q_m)=|\langle\wsi|\wpi\rangle|\leq F(\rho,\omega)
\label{t4p3}
\end{equation}
due to Eq. (\ref{pr4}). Simultaneously, we have
${\cal{F}}(p_m,q_m)\geq F(\rho,\omega)$ by the statistical
interpretation (\ref{relfid0}). Hence, the equality
${\cal{F}}(p_m,q_m)=F(\rho,\omega)$ is provided. According to the
statement of Lemma 2, the fidelity $F(\rho,\omega)$ can be varied
between values ${\rm{min}}\{p_m\}^{1/2}$ and 1 by change of the vector
$|\wpi\rangle$ and the built state $\omega$. $\blacksquare$

Using Naimark's extension, Theorem 3 can be generalized to POVM
measurement, but the state $\omega$ becomes a density operator
on the extended space. So we may collect density operators into an equivalence class
according to the condition (\ref{relfid00}). Each class can
further be decomposed with respect to the value $f$ of quantum
fidelity between two states. If quantum channel ${\cal{E}}$ is
unistochastic then the general scheme of arguments may be extended
in terms of partial fidelities. Uhlmann \cite{uhlmann00}
introduced the $k$-th partial fidelity between density operators
$\rho$ and $\omega$ by
$F_k(\rho,\omega):=\sum_{j>k}s_j(\sqrt{\rho}\sqrt{\omega})$, where
singular values should be put in the decreasing order. It
turned out that all the partial fidelities increase under
any unistochastic quantum operation \cite{rast093}. Some analog of the 
statistical interpretation (\ref{relfid0}) also holds \cite{rast093}. Using
unistochastic channel, the given POVM cannot be cloned with
$H_{\cal{B}}=0$ over each pair $\{\rho,\omega\}$ of states such
that ${\cal{F}}_k(p_m,q_m)^2<F_{\varkappa}(\rho,\omega)$, where
$\varkappa=(2n-k)k$ and $n$ denotes the number of POVM elements. We refrein from
presenting the explicit calculations.

\section{Example of perfect cloning}

We consider the simplest case when
$d=2$, the measurement is projective and the qubits `A' and `T'
interact unitarily. For brevity, we will write $|\eta\rangle$ and
$|\theta\rangle$ instead of $|\eta_{+}\rangle$ and
$|\theta_{+}\rangle$ respectively. We also define states
$|\varphi\rangle:=\sin\eta|0\rangle+e^{i\varphi}\cos\eta|1\rangle$.
The PVM measurement $\{|0\rangle\langle0|,|1\rangle\langle1|\}$ on
the state $|\eta\rangle$ generates probability distribution
$\{\cos^2\eta,\sin^2\eta\}$;  on the state $|\varphi\rangle$ the
one generates distribution $\{\sin^2\eta,\cos^2\eta\}$ independent
of $\varphi$. So, the classical fidelity between these two
distributions is ${\cal{F}}=\sin2\eta$. In general, the two
distributions differ and ${\cal{F}}<1$. We shall now describe a
procedure for perfect cloning of the PVM over pair $|\eta\rangle$
and $|\varphi\rangle$. The overlap between these states is equal
to
\begin{equation}
\langle\eta|\varphi\rangle=\cos\eta\sin\eta(1+e^{i\varphi})=e^{i\varphi/2}\sin2\eta\cos(\varphi/2)
\ . \label{tgov}
\end{equation}
The perfect cloning is possible when
$|\langle\eta|\varphi\rangle|\leq{\cal{F}}^2$, and hence
$|\cos(\varphi/2)|\leq\sin2\eta$. Further, we assume that
$\langle\eta|\varphi\rangle\not=0,1$. Indeed, orthogonal (and
identical) states can be perfectly cloned, so this case does not
add anything new into discussion. Initially, Trudy prepares her
qibit `T' in the state $|\eta\rangle$. Then she performs a unitary
transformation on ${\mathbb{C}}^2\otimes{\mathbb{C}}^2$ specified
by
\begin{equation}
{\U}{\>}|\eta\rangle\otimes|\eta\rangle=|\eta\rangle\otimes|\eta\rangle
\ , {\quad}
{\U}{\>}|\varphi\rangle\otimes|\eta\rangle=|\varphi'\rangle\otimes|\varphi''\rangle
\ . \label{ugt}
\end{equation}
By the unitarity constraint, we have
$\langle\eta|\varphi\rangle=\langle\eta|\varphi'\rangle\langle\eta|\varphi''\rangle$,
whence
$\exp(i\varphi/2)\cos(\varphi/2)=\exp\bigl[i(\varphi'+\varphi'')/2\bigr]\sin2\eta\cos(\varphi'/2)\cos(\varphi''/2)$.
The ideal case $\varphi'=\varphi$ and $\varphi''=\varphi$ is
forbidden by the no-cloning theorem for quantum states. But this
is not necessary for the exact replication of the statistics. For
arbitrary values of $\varphi'$ and $\varphi''$, the needed
distribution $\{\sin^2\eta,\cos^2\eta\}$ is provided at the output
on both the qubits `A' and `T'. The choice  $\varphi'=\varphi''$
leads to $\varphi=2\varphi'$ and
$2\cos(\varphi/2)=\sin2\eta(1+\cos\varphi')$, whence
$\cos(\varphi/2)=\cos\varphi'=\sin2\eta/(2-\sin2\eta)$. For given
$\eta$, we calculate values of $\varphi'$ and $\varphi$. [For
those values of $\varphi$ that violate the last condition, the
scheme with $\varphi'\neq\varphi''$ works.] Note that the
requirement $|\langle\eta|\varphi\rangle|\leq{\cal{F}}^2$ is
herewith provided. The states $|\eta\rangle$ and $|\theta\rangle$
form an orthonormal basis in ${\mathbb{C}}^2$. In this basis, we
write $|\varphi\rangle=\alpha|\eta\rangle+\beta|\theta\rangle$ and
$|\varphi'\rangle=\alpha'|\eta\rangle+\beta'|\theta\rangle$. Using
Eqs. (\ref{ugt}), we find a representation of transformation $\U$
as $4\times4$--matrix with respect to the basis
$\{|\eta\eta\rangle,|\eta\theta\rangle,|\theta\eta\rangle,|\theta\theta\rangle\}$.
For example, we can take
\begin{equation}
\left(\begin{array}{cccc}
 1 & 0 & 0 & 0 \\
 0 & auv^{*}|u|^{-1} & u & a|u|^2 \\
 0 & 0 & u & -a^{-1} \\
 0 & -a|u| & v & avu^{*}
\end{array}\right)
\ , \label{umatr}
\end{equation}
where $u=\alpha'\beta'\beta^{-1}$, $v=\beta'\beta'\beta^{-1}$ and
$a^2=(|u|^2+|v|^2)^{-1}$. The made calculations are outlined in
Appendix B. It can be checked that the four column vectors of this
matrix are mutually orthogonal and unit. Note that only the first and
third columns are strictly kept by the requirement (\ref{ugt}). The
second and fourth columns are varied under the unitarity property.

We have described the example of perfect cloning of single PVM.
This example is non-trivial, because
$\langle\eta|\varphi\rangle\not=0,1$ and the original probability
distributions $\{\cos^2\eta,\sin^2\eta\}$ and
$\{\sin^2\eta,\cos^2\eta\}$ are different. For each of the two
states $|\eta\rangle$ and $|\varphi\rangle$ at the input, the
measurement on both the qubits `A' and `T' at the output gives the
factorized distribution $\{t_{jk}\}=\{p_jp_k\}$. The perfect cloning 
became possible due to the validity of
$|\langle\eta|\varphi\rangle|\leq{\cal{F}}^2$. That is, the pair
$\{|\eta\rangle,|\varphi\rangle\}$ was chosen to be tolerant to
cloning of PVM $\{|0\rangle\langle0|,|1\rangle\langle1|\}$.

\section{Conclusions}
 
We have examined the notion of POVM cloning as a special
strong form of broadcasting. As figure of merit, the relative
entropy of the actual joint distribution to the factorized
distribution was proposed. In parallel with the incompatibility, a
character of potential input states can prevent the exact
replication of statistics at the output. So we approached a
problem from a somewhat different point of view than the previous
studies of observable cloning. The no-cloning theorem for a single
POVM measurement has been established. The presented reasons have
been illustrated on the example related to the B92 protocol. We have
also investigated some properties of those sets that are
intolerant to perfect cloning of given POVM. The simple example of
perfect cloning has been described explicitly. This example shows
that the cloning with factorized distribution at the output may be
real too.

\acknowledgments
The present author is grateful to anonymous referees for useful 
comments. This work was supported in a part by the Ministry of
Education and Science of the Russian Federation under grants 
no. 2.2.1.1/1483, 2.1.1/1539.

\appendix

\section{Proof of Lemma 2}

Let us suppose that
${\qp}_m=\sum\nolimits_{n=1}^{N(m)}|e_{mn}\rangle\langle e_{mn}|$,
where $N(m)$ denotes rank of projector ${\qp}_m$ and a basis
$\{|e_{mn}\rangle\}$ is orthonormal. The vectors $|\psi\rangle$
and $|\phi\rangle$ are expressed as
\begin{equation}
|\psi\rangle=\sum\nolimits_m \sum\nolimits_{n=1}^{N(m)} c_{mn}|e_{mn}\rangle
\ , {\quad}
|\phi\rangle=\sum\nolimits_m \sum\nolimits_{n=1}^{N(m)} b_{mn}|e_{mn}\rangle
\ . \label{reprh}
\end{equation}
Multiplying each $|e_m\rangle$ by phase factor properly, we can
always make all the coefficients $c_{mn}$ to be positive reals. We
also restrict attention to superposition (\ref{reprh}) with
positive real coefficients $b_{mn}$. Then the fidelity between
states is expressed as
$|\langle\psi|\phi\rangle|=\sum_{mn}c_{mn}b_{mn}$. By $p_m=\sum_n
c_{mn}^2$ and $q_m=\sum_n b_{mn}^2$, the classical fidelity is
rewritten as
\begin{equation}
{\cal{F}}(p_m,q_m)=\sum\nolimits_m \left(\sum\nolimits_{n=1}^{N(m)} c_{mn}^2\right)^{1/2}
\left(\sum\nolimits_{n=1}^{N(m)} b_{mn}^2\right)^{1/2}
\ . \label{cdeg}
\end{equation}
The equality $|\langle\psi|\phi\rangle|={\cal{F}}(p_m,q_m)$ takes
place if and only if
\begin{equation}
\sum\nolimits_{n=1}^{N(m)}  c_{mn}b_{mn}=\left(\sum\nolimits_{n=1}^{N(m)} c_{mn}^2\right)^{1/2}
\left(\sum\nolimits_{n=1}^{N(m)} b_{mn}^2\right)^{1/2}
\ . \label{casch}
\end{equation}
for all $m$. At fixed $m$, the $N$--tuples
$(c_{m1},\ldots,c_{mN})$ and $(b_{m1},\ldots,b_{mN})$ of real
numbers can be viewed as $N$--dimensional vectors of real space
${\mathbb{R}}^N$. According to the Cauchy--Schwarz inequality, the
equality (\ref{casch}) is valid if and only if these two vectors
are linearly related. In other words, for all $n$ and fixed $m$ we
have $b_{mn}=\gamma_m c_{mn}$ with some positive $\gamma_m$. Of
course, numbers $\gamma_m$ are varied under the condition
$\langle\phi|\phi\rangle=1$. The maximum
$\langle\psi|\phi\rangle=1$ is reached when $b_{mn}=c_{mn}$ for
all $m$ and $n$. We can also write
\begin{equation}
{\cal{F}}(p_m,q_m)\geq {\rm{min}}\{p_m\}^{1/2}\sum\nolimits_m \sqrt{q_m}
\geq {\rm{min}}\{p_m\}^{1/2}\sum\nolimits_m q_m={\rm{min}}\{p_m\}^{1/2}
\ . \label{fich}
\end{equation}
This lower bound can always be reached. Let $m_0$ be value such
that $p_{m_0}={\rm{min}}\{p_m\}$. We take
$b_{m_0n}=p_{m_0}^{-1/2}c_{m_0n}$ and $b_{mn}=0$ for $m\neq m_0$,
whence $q_{m_0}=1$ and $q_{m}=0$ for $m\neq m_0$. Here the
equality (\ref{casch}) still holds and
${\cal{F}}(p_m,q_m)=p_{m_0}^{1/2}$. As a continuous function of
$b_{mn}$, the quantity $|\langle\psi|\phi\rangle|$ ranges between
${\rm{min}}\{p_m\}^{1/2}$ and 1. $\blacksquare$

\section{Calculation of matrix elements}

Let us represent basis vectors $|0\rangle$ and
$|1\rangle$ as combinations of $|\eta\rangle$ and
$|\theta\rangle$. Solving the needed equations, we get
$|0\rangle=\cos\eta|\eta\rangle-\sin\eta|\theta\rangle$,
$|1\rangle=\sin\eta|\eta\rangle+\cos\eta|\theta\rangle$.
Substituting these terms, we get
$|\varphi\rangle=\alpha|\eta\rangle+\beta|\theta\rangle$ with
values
$\alpha=\left(1+e^{i\varphi}\right)\sin\eta\cos\eta=e^{i\varphi/2}\cos(\varphi/2)\sin2\eta$,
$\beta=-\sin^2\eta+e^{i\varphi}\cos^2\eta$. Replacing $\varphi$
with $\varphi'$, we obtain $\alpha'$ and $\beta'$ in formula
$|\varphi'\rangle=\alpha'|\eta\rangle+\beta'|\theta\rangle$. Note
that $\alpha=(\alpha')^2$ by the choice of $\varphi$ and $\varphi'$.
The specification (\ref{ugt}) is recast as the two conditions
${\U}{\,}|\eta\eta\rangle=|\eta\eta\rangle$ and
$\beta{\,}{\U}{\,}|\theta\eta\rangle=\alpha'\beta'|\eta\theta\rangle+
\beta'\alpha'|\theta\eta\rangle+\beta'\beta'|\theta\theta\rangle$.
Multiplying the first condition by each of four vectors
$|\eta\eta\rangle$, $|\eta\theta\rangle$, $|\theta\eta\rangle$,
$|\theta\theta\rangle$, we obtain the first column $(1,0,0,0)^{T}$
of the matrix (\ref{umatr}). In the same manner, the third column
$(0,u,u,v)^{T}$ of (\ref{umatr}) is derived from the second
condition. Adding columns $(0,0,1,0)^{T}$ and $(0,0,0,1)^{T}$, we
apply the Gram-Schmidt orthonormalization process which results in
the matrix (\ref{umatr}).

\end{document}